\documentclass[twocolumn,prl,superscriptaddress,floatfix,showpacs]{revtex4-1}
\usepackage[utf8]{inputenc}
\usepackage{amsmath}
\usepackage{amssymb}
\usepackage{graphicx}

\makeatletter
\usepackage{graphics}
\usepackage{epsfig}
\usepackage{color}

\newcommand{\be}{\begin{equation}} \newcommand{\ee}{\end{equation}}
\newcommand{\bea}{\begin{eqnarray}} \newcommand{\eea}{\end{eqnarray}}

\makeatother

\begin{document}

\title{Quantifying the dynamical complexity of time series}

\author{Antonio Politi} 
\affiliation{Institute for Complex Systems and Mathematical Biology, SUPA, University of Aberdeen, Aberdeen, UK}

\date{\today}

\begin{abstract}
A powerful tool is developed for the characterization of chaotic signals.
The approach is based on the symbolic encoding of time series (according to their ordinal
patterns) combined with the ensuing characterization of the corresponding cylinder sets.
Quantitative estimates of the Kolmogoro-Sinai entropy are obtained by introducing a modified
permutation entropy which takes into account the average width 
of the cylinder sets. The method works also in hyperchaotic systems and allows estimating
the fractal dimension of the underlying attractors.
\end{abstract}

\maketitle

Since the discovery of deterministic chaos, the problem of distinguishing irregular deterministic from
stochastic dynamics has attracted the interest of many scientists who have thereby proposed different approaches.
In principle the Kolmogorov-Sinai (KS) entropy $h_{KS}$ is the most appropriate indicator:
it quantifies the growth rate of the number of distinct trajectories generated by a given dynamical
system, when the length of the trajectories is increased~\cite{sinai09}.
In stochastic processes $h_{KS}$ is infinite, while in deterministic chaotic systems,
the Pesin formula implies that $h_{KS}$ is smaller than or equal to the sum of the positive Lyapunov
exponents (LEs)~\cite{pesin77}.
Unfortunately, it is difficult to obtain directly reliable estimates of $h_{KS}$. 
Its computation requires partitioning the phase space into cells (the atoms), 
so that any trajectory can be encoded as a suitable symbolic sequence. 
However, only generating partitions ensure a correct estimate of $h_{KS}$~\cite{eckmann85}:
generic partitions give lower bounds, whose quality is a priori unknown. 
Effective procedures to construct generating partions have been developed at most for two-dimensional maps 
(or, equivalently, for three-dimensional continuous-time attractors).
They are based on the so-called primary homoclinic tangencies which have to be connected in a suitable
order~\cite{grassberger85} (when the dynamics is dissipative) and symmetry lines which allow 
splitting the stability islands~\cite{christiansen97} (when the dynamics is Hamiltonian). 
In any case the procedure requires much work, including an accurate identification of
the locally stable and unstable manifolds. Even worse, extensions to higher dimensions are not available.

Alternative approaches have been proposed, based on various types of symbolic encoding
(see, e.g.,~\cite{kennel03,hirata04,buhl05}), none of which, goes, however, beyond two-dimensional maps.
A particularly appealing method was proposed by Bandt and Pompe~\cite{bandt02a}, who proposed to look at 
the relative ordering of sequentially sampled time series~\cite{bandt02a}. 
The growth rate $k_P$ of the corresponding {\it permutation entropy} can be easily computed and is
often used as a proxy for 
$h_{KS}$. The advantage of this approach is that symbolic sequences are obtained without the need 
of explicitly partitioning the phase space and can thus be used as a zero-knowledge approach 
for the analysis of experimental time series. In fact, the permutation entropy has been widely
used in many different contexts: see, e.g.~\cite{cao04,staniek07,zanin12,masoller15,weck15}.
In 1d and 1d-like maps, it has been proved that $k_P$ is equal to $h_{KS}$~\cite{bandt02b,amigo05}. 
Unfortunately, even when $k_P$ is expected to coincide with $h_{KS}$, it is affected by so strong
finite-size corrections as to make extrapolations questionable. Given such difficulties, 
some researchers have proposed to use suitable combinations of different indicators to provide a sharper
characterization of chaotic signals~\cite{mancini95,rosso07}.  
However, such strategies do not go that far, as they basically rely on the same background information: the proability
of the different symbolic sequences.

In this Letter, I show that substantial progress can be made by including in the analysis 
the ``dispersion'' among trajectories characterized by the same symbolic sequence. 
This information, which has been so far overlooked, does not only contribute to a better discrimination
between stochastic and deterministic dynamics, but allows also obtaining quantitative estimates 
of $h_{KS}$ even in dynamical systems characterized by more than one positive LE.

{\it Models}: Before starting the theoretical considerations and the numerical analysis, I 
introduce the four dynamical systems of increasing complexity used here as a testbed:
(i) the H\'enon map, for the usual parameter values ($z(t+1) = a - z(t)^2 + b z(t-1)$, with  
$a=1.4$ and $b=0.3$); (ii) one of the various versions of the R\"ossler model 
($\dot z_1 = z_2-z_3$; $\dot z_2 = z_1 + az_2$;  $\dot z_3 = b + z_3(z_1-c)$, with $a=1/2$, $b=2$, 
$c=4$); (iii) a generalized 3d H\'enon map (GH) ($z(t+1) = a - z(t-1)^2-bz(t-2)$, with $a=1.5$ and 
$b=0.29$)~\cite{note0};
(iv) the Mackey-Glass model ($\dot z = 2z(t-t_d)/[1+z(t-t_d)^{10}] - z$, with $t_d=3.3$).

\begin{figure}
\begin{centering}
\includegraphics[width=0.4\textwidth,clip=true]{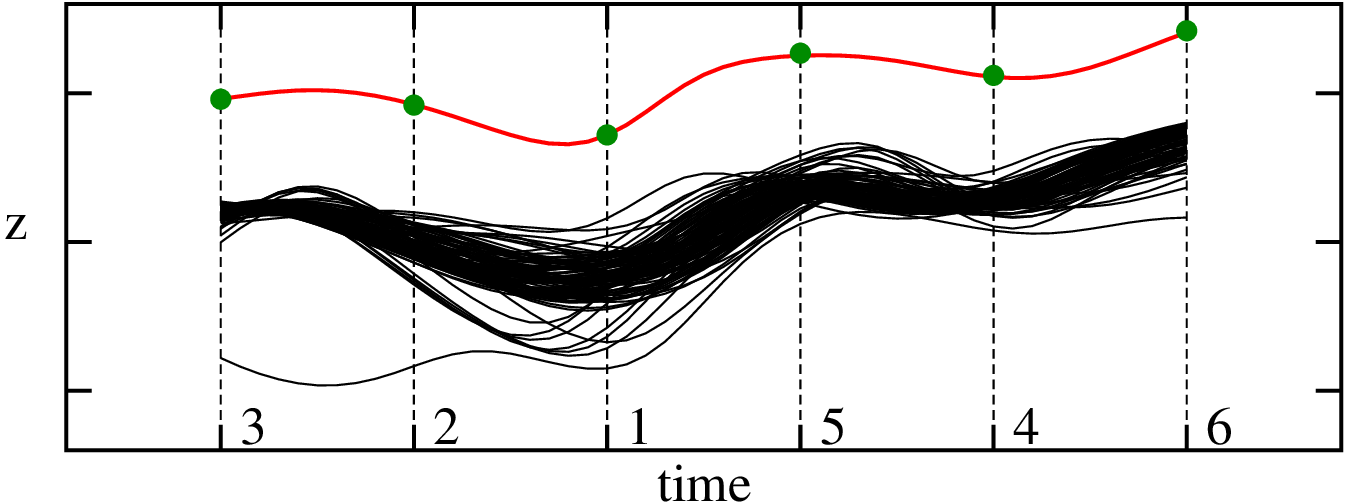}
\end{centering}
\caption{\label{fig:delta}
100 trajectories of the Mackey-Glass model characterized by the same ordinal pattern
(3,2,1,5,4,6) (see the text for a detailed explanation).
The dashed vertical lines identify the sampling times, when the signal is measured. 
One of the trjectories is arbitrarily shifted to better show the ordering of
the sampled points (see the green dots).}
\end{figure}

I start recalling the concept of permutation entropy and introducing the relevant observables. 
Let $z(t)$ denote a scalar variable that is assumed to be sampled every $T$ time units, 
so that $z_n = z(t= T n)$.  Next, consider a moving time window of length $m$.
An example is given in Fig.~\ref{fig:delta}: the top red curve is a piece of trajectory of
the Mackey-Glass system, while the green dots 
represent the sampled points (a window of length $m=6$ is there considered, with $T=1$). 
Such a specific sequence is encoded as (3,2,1,5,4,6): each single number denotes the 
ordinal position (from the smallest to the largest point within the window itself), 
so that the ``1" in position 3 means that the third element within the window is the smallest one. 
The trajectories can be grouped according to their symbolic sequence. 
Given the probabilities $p_i$ of all the sequences of length $m$, 
one can determine the ``permutation entropy" $K_P(m) = - \langle \log p_i \rangle$.
Its growth rate $k_P = (K_P(m+1)-K_P(m))/T$ is often used as a proxy for $h_{KS}$.

\begin{figure}
\begin{centering}
\includegraphics[width=0.45\textwidth,clip=true]{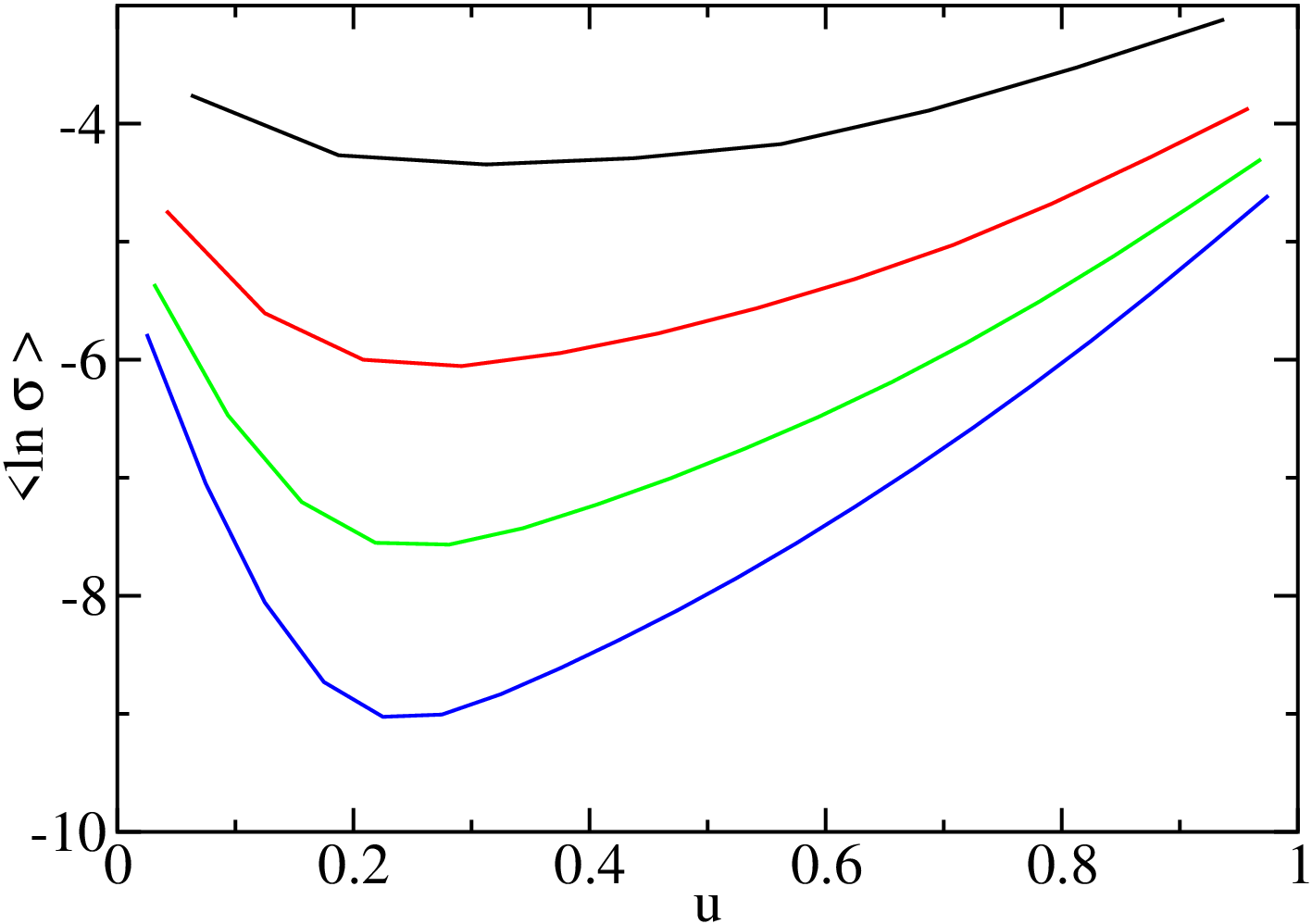}
\end{centering}
\caption{\label{fig:width2} Average cylinder width of the H\'enon maps for $m=8$, 12, 16 and 20 
(from top to bottom).  The window length is rescaled to allow for a clearer comparison ($u=(j-1/2)/m$)}
\end{figure}

Let me now focus on trajectories that are compatible with the same $i$th symbolic sequence
In Fig.~\ref{fig:delta}, I have plotted 100 randomly sampled elements, all encoded as
(3,2,1,5,4,6) (the scale is irrelevant). They are basically grouped around a ``cylinder''
of variable width: the dispersion of $z_i(j)$ ($1\le j\le m$) measures the uncertainty in the
position $j$, for the given $i$th symbolic sequence.
Here below I show that the information on the $z$-dispersion can be profitably used to characterize
a time series. There are different ways to quantify the dispersion. 
I propose to use the standard deviation $\sigma_i(j)$ of the variable $z$ measured 
in the $j$th position of the window corresponding to the $i$th sequence. Simulations
show that $\sigma_i(j)$ strongly varies with $i$; it is therefore necessary to
to average this observable over the elements of the partition: for
reasons that will become clear later, I propose to average its logarithm.
The results for the H\'enon map are reported in Fig.~\ref{fig:width2} for different values
of the window length $m$ (after rescaling $m$ to unit length). 
The average $\langle \ln \sigma_i(j)\rangle$ is computed by
weighting each sequence according to its probability $p_i$. As one can see, it varies
along the window and, more important, it decreases upon increasing $m$. 

\begin{figure}
\begin{centering}
\includegraphics[width=0.4\textwidth,clip=true]{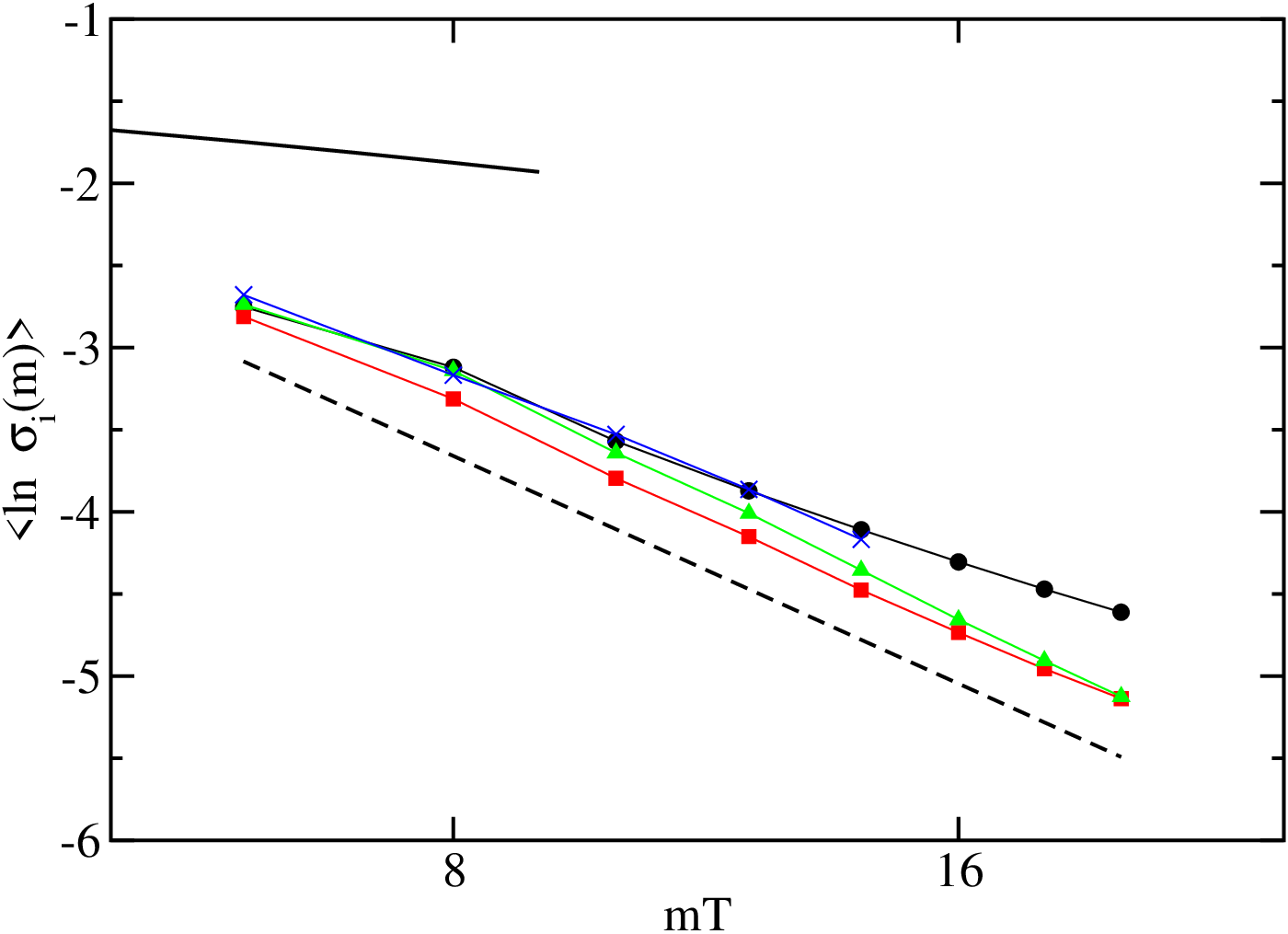}
\end{centering}
\caption{\label{fig:distance} The average logarithm of the dispersion $\sigma_i(m)$ in the last element
of a window of length $m$ for various dynamical systems. Circles, squares, triangles and crosses
refer to the H\'enon map, R\"ossler attractor, GH map, and Mackey-Glass, respectively.
The upper solid curve corresponds to a linear stochastic process (see the text), while the dashed line
illustrates a $1/m^2$ decrease.}
\end{figure}

This is not accidental. In Fig.~\ref{fig:distance} I plot the dependence of $\langle \ln \sigma_i(m)\rangle$
on the window length for the above mentioned dynamical systems. Straight lines correspond to 
a power-law decrease of $\sigma$. The dashed line, drawn for reference, corresponds to a $1/m^2$ dependence,
a behavior approximately followed in all of the deterministic models.

For the sake of comparison, I have also plotted the results for a stochastic signal: 
the linear process $z(t+1)= \gamma z(t) +\xi(t)$, where $\xi(t)$ is a random number uniformly
distributed within $[-0.5,0.5]$ and $\gamma=1/2$ (see the upper solid line, whose slope is smaller than 1/2).
This figure provides a first evidence of the usefulness of the cylinder width $\sigma$:
it allows distinguishing deterministic, from stochastic signals. However, in this Letter rather than further
exploring this point I wish to focus on a different application of $\sigma$,
which makes it possible to obtain accurate estimates of $h_{KS}$.

In this perspective, it is necessary to recall the definition of Kolmogorov-Sinai 
entropy. Consider an $N$-dimensional variable ${\bf x}(t)$ which evolves in time and denote with 
${\bf C}_i$ a cylinder in $\mathbb{R}^N\times \mathbb{R}$ 
of width $\varepsilon_i$ and time length $\tau$, centered around some trajectory
${\bf x}(t')$ for $t<t'<t+\tau$; let also $p_i(\varepsilon_i,\tau)$ denote the probability
that a generic trajectory of length $\tau$ is fully contained in ${\bf C}_i$.
By then covering $\mathbb{R}^N\times [t,t+\tau]$ with non-overlapping cylinders of
width $\varepsilon_i$ (for reasons that will become clear later on, I assume that the various
cylinders may have different widths), one can introduce the entropy
$ H(\varepsilon,\tau) =  - \langle p_i \ln p_i(\varepsilon_i,\tau)\rangle$ ($\varepsilon$
without subscripts denotes a  generic average width) and thereby the Kolmogorov-Sinai entropy as
\begin{equation}
h_{KS} =  -\lim_{\varepsilon\to 0} \lim_{\tau\to\infty} \frac{H(\varepsilon,\tau)}{\tau} \; ,
\label{eq:HK-def}
\end{equation} 
where the infinite-time limit is to be taken first. The $\varepsilon \to 0$ limit is needed to 
avoid underestimations of $h_{KS}$ (as mentioned in the introduction, this is not required in the case of 
generating partitions). From a computational point of view, it is convenient to determine the
{\it derivative}
\[
h = [H(\varepsilon,\tau+\Delta)-H(\varepsilon,\tau)]/\Delta
\]
as it converges faster than $H/\tau$ for increasing $\tau$.

The (Pesin) relationship between $h_{KS}$ and the Lyapunov exponents is based on the formula 
(see e.g. \cite{pikovsky16})
\[
p_i(\varepsilon_i,\tau) \approx \varepsilon_i^{D_i} \exp[-\Lambda_i \tau]
\]
where $D_i$ and $\Lambda_i$ are the fractal dimension and the sum of the positive
finite-time Lyapunov exponents associated with the $i$th symbolic sequence~\cite{note1}.
Upon averaging over all cylinders, one obtains
\begin{equation}
H(\varepsilon,\tau) = -D \langle \ln \varepsilon_i \rangle + \Lambda \tau
\label{eq:HD}
\end{equation}
where $D$ is the information dimension and $\Lambda$ 
the sum of the usual positive Lyapunov exponents. 
The presence of the LE implies that the limit $\varepsilon \to 0$ has been taken, since, by
definition, they refer to infinitesimal perturbations.
Eq.~(\ref{eq:HD}) implies that $h = \Lambda$, which is nothing but Pesin formula.

By now going back to the permutation entropy, we can identify $H(\varepsilon,\tau)$ with
$K_P(\sigma,mT)$, provided that a meaningful mapping between $\varepsilon_i$ and $\sigma_i$
is established. Rigorously speaking, $\varepsilon_i$ is determined in the original 
$N$-dimensional phase space, while
$\sigma_i$ refers to the measured, scalar, variable. However, we can safely identify the
two observables, since the embedding theorem proved by Takens~\cite{takens81}, ensures 
the existence of a one-to-one mapping between the original and the embedding variables
(at least when the window length $m$ is sufficiently large).
In the context of the permutation entropy, the widths $\sigma_i$ are  not given a priori, but 
self-determined by the ordering procedure and depend on the symbolic sequence: this
is the reason why, from the very beginning, I have introduced a subscript to denote the width $\varepsilon_i$.
In order to complete the identification of $\varepsilon_i$ with $\sigma_i(j)$, one should notice
that the latter indicator depends on $j$, i.e. on the position along the window
where it is determined. The theoretical argument invoked to derive the Pesin formula requires
that $\varepsilon_i$ is the maximal distance between two trajectories characterized by the same symbolic sequence.
Accordingly, I propose the identification of $\sigma_i(m)$ with $\varepsilon_i$, since in all cases I
have investigated the maximal cylinder width is attained in the last $m$th position~\cite{note2}.

The most important property of $\sigma_i(m)$ is that it decreases upon increasing the window
length (look back at Fig.~\ref{fig:distance}). This implies that $\varepsilon_i$ in Eq.~(\ref{eq:HD}) does 
depend on $\tau$ (or, equivalently, on $m$) and this invalidates the direct connection between $k_P$ and 
$\Lambda$ (and thereby with $h_{KS}$). 
A clean relationship can be re-established by introducing the relative permutation entropy
\begin{equation}
\tilde K_P(m) = K_P(m) + D \langle \log \sigma_i (m) \rangle \; ,
\label{eq:tildeK}
\end{equation}
where I have made it explicit that $\sigma$ depends on time. The derivative
$\tilde k_P = [\tilde K_P(m+1)-\tilde K_P(m)]/T$, is cleansed of the spurious time dependence 
of the cylinder-widths and thus provides a reliable estimate of $h_{KS}$.
Notice that the structure of Eq.~(\ref{eq:tildeK}) justifies the choice of averaging the
logarithm of $\sigma_i$.

{\bf Validation}: I have tested the above theoretical considerations in four different models.
The results are summarized in Fig.~{\ref{fig:all}. I start from the H\'enon map, whose analysis is 
plotted in panel (a). In this case, there is only one positive LE, $\lambda_1 \approx 0.4192$, so that
$h_{KS}$ coincides with $\lambda_1$. The fractal (information) dimension $D$, is equal to $1.258\ldots$, as
as obtained from the Kaplan-Yorke formula. The derivatives $k_P$ and $\tilde k_P$ reported in Fig.~\ref{fig:all}a 
are both determined by referring to a time interval $\Delta$ equal to 2. 
The results indicate that $\tilde k_P$, provides relatively accurate
estimates already for $\tau=9$ (here $m=\tau$).
The second model I have studied is the R\"ossler attractor, selecting the parameter values
in such a way that the dynamics is not phase-coherent, to make the attractor as different as possible 
from that of the H\'enon map. Since time is continuous, it is necessary to fix the sampling interval $T$. 
I have chosen $T=1$, which is about 1/8th of the main periodicity, but allows for an appreciable 
variation of $z_1$ (up to 1/5 of its whole range). In this case, there is again only one positive 
LE ($\lambda_1 \approx 0.1208$), but also a vanishing one, which does not contribute to the KS-entropy,
but indirectly to the correction term, affecting the dimension which is $D \approx 2.05$.
The results are plotted in Fig.~{\ref{fig:all}b. The red curve and the blue circles have been
obtained by sampling $z_1(t)$, while the green curve and the stars correspond to $z_3(t)$. 
The mutual agreement proves the robustness of the approach: there is no problem of variable selection.
Once again, the asymptotic value of the KS-entropy is achieved for $m=\tau=10$, 
where the permutation-entropy estimates are still five times larger than expected.

\begin{figure}
\begin{centering}
\includegraphics[width=0.45\textwidth,clip=true]{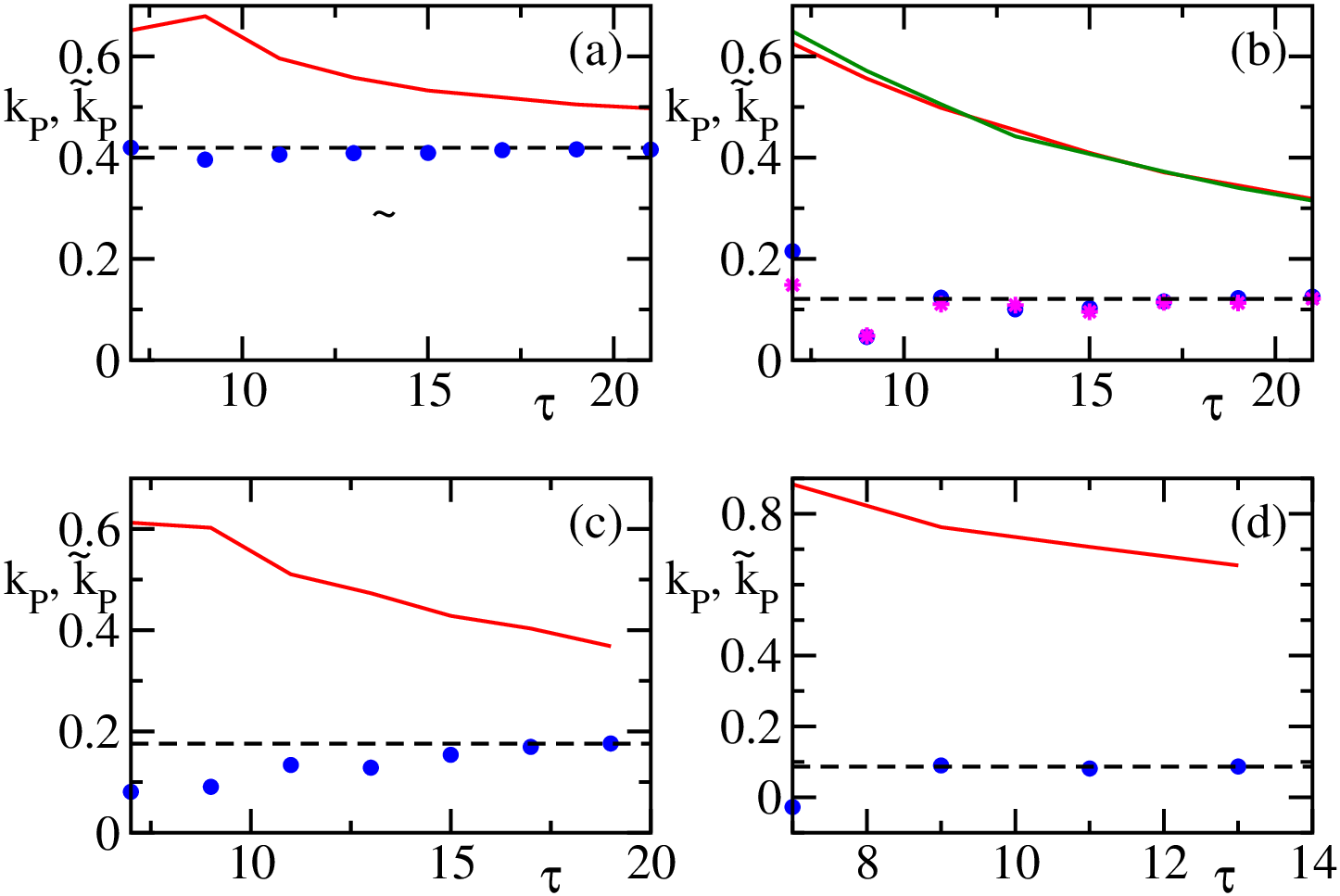}
\end{centering}
\caption{\label{fig:all}Finite-size estimates of the Kolmogorov Sinai entropy. 
The solid curves correspond to the derivative $k_P$ of the permutation entropy; 
symbols to the derivative $\tilde k_P$ of the relative entropy~\ref{eq:tildeK}.
The four panels report the results for different models:
(a) H\'enon map, (b) R\"ossler attractor, (c) GH map; (d) Mackey-Glass model.
In all cases the horizonatal black dashed line corresponds to the KS-entropy estimated as the
sum of the positive Lyapunov exponents.}
\end{figure}

Next, I turned my attention to another discrete-time system, the GH map, where there are
two positive Lyapunov exponents ($\lambda_1 \approx  0.1179$, $\lambda_2 \approx 0.0577$), so that the 
KS-entropy is $0.1756\ldots$.
while the dimension is $D\approx 2.12$. Also in this case, $\tilde k_P$ exhibits a convicing agreement 
with the theoretical expectations. Being scrupulous, for $m=13$ there are still deviations 
of order $25\%$ (to be, however, compared with the 6-times larger overestimation of the traditional method). 
Finally, I have studied the Mackey-Glass equation.
This is a model with delayed interactions, i.e. the phase-space is infinite dimensional. For the parameter
value I have selected there are two positive LEs ($\lambda_1 \approx 0.0617$, $\lambda_2 \approx 0.0234$) so that
$h_{KS} \approx 0.0862$, and the Kaplan-Yorke dimension is $D \approx 3.73$. By looking at Fig.~{\ref{fig:all}d, one
can see that a remarkable agreement is found already for $\tau = 9$ (also in this case the sampling
time $T$ has been fixed equal to 1).

Altogether, the time-dependence of the cylinder width helps to resolve an ostensible paradox: finite partitions
typically tend to underestimate the KS-entropy, because they are unable to discriminate all 
different trajectories. Nevertheless, the permutation entropy overestimates $h_{KS}$:
this is because part of the entropy increase is a spurious effect induced by the implicit refinement of the phase-space 
partition. The modified definition herein proposed gets rid of such a contribution.

The dependence of $\langle \log \sigma_i(m) \rangle$ on $m$ reported in Fig.~\ref{fig:distance}
shows that $\sigma$ decreases as a power law, $\sigma\approx m^{-\gamma}$ with $\gamma$ close to 2.
By combining this observation with the assumption that this is the major source of finite-size corrections
(at least in a suitable range of $m$ values), one can claim that $\tilde K_P(m) = K_0+ \Lambda m$ and
thereby write
\[
K_P(m) = K_0 + \Lambda m + D \gamma \ln m  \; .
\]
This equation, at the same time, suggests that the derivative itself of the standard permutation entropy 
eventually converges to $\Lambda$, but also that it is affected by strong (logarithmic in the window length)
corrections. They make the estimation of the asymptotic value prohibitive.

In this Letter I have shown that reliable estimates of $h_{KS}$ are obtained 
without the need of explicitly partitioning the phase space,
but this requires the knowledge of the fractal dimension $D$. Now I show that this obstacle
can be overcome when the major source of finite-size corrections is the above mentioned
logarithmic term. With reference to  Eq.~(\ref{eq:tildeK}), I replace $D$ with an  unknown parameter $d$
and thereby introduce $\tilde K_P(m,d)$ ($\tilde K_P(m,0)$ coincides with the usual permutation entropy). 
So long as $d<D$, the derivative $\tilde k_P(m,d)$, converges to the asymptotic value from above,
while a convergence from below is expected when when $d>D$. Therefore, one can hope to estimate $D$ as
the critical $d$-value such that  $\tilde k_P(m,d)$ is independent of $m$.
With this idea in mind, one can go to the initial data and determine $\tilde k_P(m,d)$ in a suitable range of $m$ 
values for different $d$-values.
A linear fit of the last seven points for the H\'enon and Roessler attractors shows that the average derivative
of $\tilde k(m,d)$ changes sign for $d\approx 1.13$ and $d \approx 1.9$, respectively. 
For the GH map, the change of sign occurs  for $d \approx 1.56$, while for the Mackey Glass model, I obtain $d \approx 3.8$. 
All values are close to the expected estimates of the dimension, with the exception
of the generalized H\'enon map, whose dimension is underestimated by about 0.6.
This is understandable, since from Fig.~\ref{fig:all}c one can see that such a dynamical system is 
the only one where the convergence of $\tilde k_P$ is not perfect. 
The reason is probably due to a slow convergence of the dimension itself to its asymptotic
value: in other words, it is reasonable to interpret the value $d=1.56$ as the {\it effective} dimension of
the attractor on the scales that are acccessed by the numerical analysis.
A more detailed study is, however, required to validate this conjecture

Altogether, I have shown that the dispersion of trajectories characterized by
the same ordinal sequencies contains important information which helps to estimate
the Kolmogorov-Sinai entropy through a modified permutation entropy even in the case
multiple positive Lyapunov exponents. The same approach can be used
as a zero-knowledge tool to determine the effective dimension over the accessible
resolution scales. All of these results are possible because the increase of
the window length corresponds to the simultaneous increase of both
the embedding dimension~\cite{kantz04} and of the resolution in phase space.
I am confident that this method can be profitably extended to mixed signals characterized
by a combination of determinism and randomness. The scaling behaviour of the
cylinder width reported in Fig.~\ref{fig:distance} represents a first encouraging
step in that direction.

\acknowledgments
The author wishes to acknowledge G. Giacomelli, M. Mulansky, and L. Ricci for early dicussions.



\begin{thebibliography}{99}
\bibitem{note0} Initial conditions which lie in the (small) basin of attraction are: $z(0)=-0.07$, 
$z(1)=0$, $z(2)=0.07$.
\bibitem{note1} Generic systems are ``multifractal", i.e. are characterized by a spectrum of
different dimensions and LEs.
\bibitem{note2} Simulations of other models suggest that this is always true in deterministic models,
while no position dependence is found in purely stochastic systems.
\end{thebibliography}
\end{document}